# Can education correct appearance discrimination in the labor market?

**Hambur Wang**

**Guanghua School of Management, Peking University, Beijing 100871, China**

## Abstract

This study explores the impact of appearance discrimination in the labor market and whether education can mitigate this issue. A statistical analysis of approximately 1.058 million job advertisements in China from 2008 to 2010 found that about 7.7% and 2.6% of companies had explicit requirements regarding candidates' appearance and height, particularly in positions with lower educational requirements. Literature review indicates that attractive job seekers typically enjoy higher employment opportunities and wages, while unattractive individuals face significant income penalties. Regression analysis of 1,260 participants reveals a significant positive correlation between attractiveness scores and wages, especially in low-education groups. Conversely, in high-education groups, the influence of appearance on income is not significant. The study suggests that enhancing education levels can effectively alleviate income declines associated with appearance, providing policy recommendations to reduce appearance discrimination in the labor market.

## 1.Research Background

In contemporary society, the combination of an individual's appearance, physique, features, and skin color is often summarized as "attractiveness." The "2018 China Youth Attractiveness Competitiveness Report" indicates that job seekers generally believe that appearance impacts employment opportunities, ranking just below work experience and educational background. The report reveals that about 70% of professionals are willing to spend over 20% of their salary to enhance their attractiveness. Additionally, 70.6% of employers stated that the decision to select candidates based on appearance depends on the specific requirements of the industry and position.

Furthermore, a statistical analysis of approximately 1.058 million job advertisements in China from 2008 to 2010 found that 7.7% and 2.6% of companies had explicit requirements regarding candidates' appearance and height. Notably, for companies with lower degree requirements, the proportions of those with explicit appearance and height requirements reached 15% and 9.3%, respectively. This suggests that attractiveness not only influences job opportunities for applicants but also affects salaries and career development. It can be argued that physical attractiveness can rival or even surpass human capital in importance.

In recent years, the term "attractiveness" has remained a frequently used keyword on the internet, indicating that society has increasingly emphasized appearance, leading to a "looks-oriented" era. Therefore, studying the impact of appearance on wage income is crucial for improving employment practices and enhancing social equity.

Currently, research on appearance discrimination by scholars both domestically and internationally primarily approaches the issue from the perspective of labor economics, attributing appearance discrimination in the labor market to employer and consumer biases. In terms of career choice, individuals with high attractiveness tend to actively select occupations that highlight their appearance, thereby gaining preference from both employers and consumers. Conversely, those with average or below-average looks often avoid careers that require attractive appearances, resulting in a natural and objective wage disparity.

Research on the impact of appearance on individual employment, wages, and career development has been conducted by foreign scholars for a longer time, becoming more mature and systematic than that of domestic researchers. Hamermesh and Biddle were the first to empirically examine the phenomenon of the "beauty premium," where attractive individuals typically earn above-average incomes, while those deemed unattractive face a "beauty penalty" with earnings below the average. They noted that both the "beauty penalty" and "beauty premium" are more pronounced for men than for women, summarizing their findings in subsequent works as "the economics of beauty."[5-7].Individuals with attractive appearances have advantages both in terms of hiring opportunities and initial salary levels[8].Marlowe and Schneider found that appearance discrimination is particularly pronounced in the success of male job seekers.[9].Hosoda and colleagues pointed out that attractive individuals generally fare better in the job market than those with average looks, and this form of appearance discrimination is equally significant for both genders[10].

Domestic scholars have confirmed the existence of varying degrees of appearance and body discrimination in China's labor market. Job seekers with attractive appearances are often more likely to receive higher evaluations compared to those with less appealing looks, and there is a close relationship between appearance and personal income [11]. Liu Yipeng argues that while the "beauty premium" does not exist, the "beauty penalty" is significant [13]. Wang Xun reached an opposing conclusion, finding that there is a "beauty premium" compared to average appearance, but no "beauty penalty," and he noted that health has a corrective effect on the relationship between appearance and income[14].Guo Jiqiang discovered that the relationship between appearance and income follows a "high-heeled shoe curve," indicating that while being attractive generally helps to increase income, the "beauty premium" for the most attractive individuals is lower than that for those who are merely above average in attractiveness. In terms of gender, Yuan Huijuan and Zhang Zhiyong found that male respondents place greater importance on appearance in hiring decisions, suggesting that men's appearance bias is more pronounced than that of women [16]. Yang Yuanzheng and colleagues argued that both "beauty premium" and "beauty penalty" exist among male and female groups, but the impact of appearance on the wages of highly educated workers is not significant[17].

## 2.Research Methodology

## 2.1. Descriptive Statistics of the Sample

This study utilizes relevant data from "WAGE," comprising a sample of 1,260 individuals, including 436 females and 824 males. Table 2-1 presents the descriptive statistics for the relevant variables.

The descriptive results in Table 2-1 indicate that the average attractiveness score for the entire sample is 3.19. The average attractiveness score for females is 3.20, while for males it is slightly lower at 3.18. The sample is categorized into a high-education group (550 individuals) and a low-education group (710 individuals) based on whether their years of education are 13 or more. The mean attractiveness scores for these groups are 3.26 and 3.13, respectively. Following Liu Yipeng's [13] methodology, the average attractiveness score is used to delineate three categories based on a scale of 1 to 5: scores of 1-2 indicate low attractiveness (155 individuals), a score of 3 indicates average attractiveness (722 individuals), and scores of 4-5 indicate high attractiveness (383 individuals).

The overall average hourly wage for the sample is 6.307 yuan, with females earning an average hourly wage of 4.30 yuan and males earning 7.37 yuan, which is higher than that of females. The average hourly wages for individuals categorized by attractiveness are 6.27 yuan for those with low attractiveness, 6.51 yuan for those with average attractiveness, and 6.35 yuan for those with high attractiveness. The average hourly wages for the high-education and low-education groups are 5.55 yuan and 7.29 yuan, respectively.

Table 2-1: Descriptive Statistics of the Sample

|  | Overall | Female | Male | Low Attractiveness | Average Attractiveness | High Attractiveness | Low Education | High Education |
|---|---|---|---|---|---|---|---|---|
| Wage (yuan/hour) | 6.307 | 4.299 | 7.369 | 5.269 | 6.505 | 6.353 | 5.549 | 7.285 |
| Work Experience (years) | 18.206 | 14.142 | 20.357 | 19.600 | 19.492 | 15.219 | 19.521 | 16.509 |
| Attractiveness Score | 3.186 | 3.202 | 3.177 |  |  |  | 3.128 | 3.260 |
| Association Membership | 0.272 | 0.211 | 0.305 | 0.232 | 0.310 | 0.217 | 0.318 | 0.213 |
| Physical Health | 0.933 | 0.920 | 0.941 | 0.929 | 0.921 | 0.958 | 0.921 | 0.949 |
| Black | 0.074 | 0.115 | 0.052 | 0.045 | 0.080 | 0.073 | 0.083 | 0.062 |
| Female | 0.346 |  |  | 0.381 | 0.323 | 0.376 | 0.346 | 0.345 |
| Married | 0.691 | 0.489 | 0.799 | 0.703 | 0.716 | 0.640 | 0.717 | 0.658 |
| Residing in Southern Region | 0.175 | 0.163 | 0.181 | 0.206 | 0.163 | 0.183 | 0.128 | 0.235 |
| Living in Large City | 0.219 | 0.204 | 0.227 | 0.200 | 0.224 | 0.217 | 0.185 | 0.264 |
| Living in Small City | 0.467 | 0.468 | 0.466 | 0.419 | 0.486 | 0.449 | 0.448 | 0.491 |
| Engaged in Service Industry | 0.274 | 0.438 | 0.187 | 0.303 | 0.255 | 0.298 | 0.172 | 0.405 |
| Years of Education | 12.563 | 12.596 | 12.546 | 11.929 | 12.421 | 13.089 |  |  |

Note: The attractiveness score ranges from 1 to 5. The variables "Association Membership," "Physical Health," "Black," "Female," "Married," "Residing in Southern Region," "Living in Large City," "Living in Small City," and "Engaged in Service Industry" are all binary variables, coded as 1 for "Yes" and 0 for "No."

## 2.2. Method Overview

This study employs a stepwise regression method to initially identify the explanatory variables required for the model. Additionally, the RESET test is conducted to further determine whether any important nonlinear relationships among the explanatory variables have been omitted. The analysis compares different forms of the dependent variable, specifically wages and the logarithm of wages, to ascertain which form is more appropriate for inclusion in the model regression. Furthermore, the Breusch-Pagan (BP) test and White test are utilized to assess the presence of heteroscedasticity in the model.

# 3.Research Process

## 3.1. Model Specification Issues

### 3.1.1. Control Variables

In addition to the dependent variable (wage) and core explanatory variables (attractiveness score and years of education), it is essential to consider the inclusion of other control variables to better explain the factors influencing wages. To mitigate the risk of multicollinearity caused by including an excessive number of explanatory variables, this study employs a stepwise regression method to ultimately identify all necessary explanatory variables for the model. The regression results are presented in Table 2-2.

$$wage = \beta_1 looks + \beta_2 educ + \beta_3 exper + \beta_4 exper^2 + \beta_5 female + \beta_6 bigcity + \beta_7 smllcity + \beta_8 union$$

Table 3-1: Regression Results for Different Combinations of Explanatory Variables（OLS）

|  | (1) | (2) | (3) | (4) | (5) | (6) | (7) | (8) | (9) |
|---|---|---|---|---|---|---|---|---|---|
|  | wage | wage | wage | wage | wage | wage | wage | wage | wage |
| Attractiveness Score | 0.156 | 0.413* | 0.437* | 0.405* | 0.414* | 0.429* | 0.437* | 0.430* | 0.438* |
|  | （0.190） | (0.183) | (0.182) | (0.175) | (0.175) | (0.173) | (0.173) | (0.173) | (0.173) |
| Years of Education | 0.371*** | 0.457*** | 0.423*** | 0.410*** | 0.412*** | 0.362*** | 0.373*** | 0.361*** | 0.372*** |
|  | （0.050） | (0.048) | (0.048) | (0.046) | (0.046) | (0.047) | (0.047) | (0.047) | (0.047) |
| Work Experience (years) | - | 0.114*** | 0.299*** | 0.257*** | 0.243*** | 0.247*** | 0.243*** | 0.247*** | 0.243*** |
|  | - | (0.011) | (0.040) | (0.038) | (0.039) | (0.039) | (0.039) | (0.039) | (0.039) |
| Square of Work Experience | - | - | -0.004*** | -0.004*** | -0.004*** | -0.004*** | -0.004*** | -0.004*** | -0.004*** |
|  | - | - | (0.001) | (0.001) | (0.001) | (0.001) | (0.001) | (0.001) | (0.001) |
| Female | - | - | - | -2.503*** | -2.377*** | -2.317*** | -2.271*** | -2.314*** | -2.268*** |
|  | - | - | - | (0.255) | (0.265) | (0.262) | (0.262) | (0.263) | (0.263) |
| Married | - | - | - | - | 0.488 | 0.610* | 0.603* | 0.607* | 0.599* |
|  | - | - | - | - | (0.277) | (0.275) | (0.274) | (0.276) | (0.276) |
| Living in Large City | - | - | - | - | - | 1.886*** | 1.810*** | 1.890*** | 1.816*** |
|  |  |  |  |  |  | (0.330) | (0.331) | (0.333) | (0.334) |
| Living in Small City | - | - | - | - | - | 0.661* | 0.635* | 0.661* | 0.635* |
|  |  |  |  |  |  | (0.270) | (0.270) | (0.271) | (0.270) |
| Association Membership | - | - | - | - | - | - | 0.607* | - | 0.608* |
|  |  |  |  |  |  |  | (0.264) |  | (0.264) |
| Residing in Southern Region | - | - | - | - | - | - | - | - | -0.073 |
|  |  |  |  |  |  |  |  |  | (0.458) |
| Black | - | - | - | - | - | - | - | -0.050 |  |
|  | - | - | - | - | - | - | - | (0.458) | - |
| Constant Term | 1.151 | -2.822** | -3.796*** | -2.078* | -2.375** | -2.608** | -2.878** | -2.596** | -2.860** |
|  | （0.808） | (0.857) | (0.873) | (0.859) | (0.875) | (0.873) | (0.879) | (0.881) | (0.887) |
| R-squared | 0.046 | 0.127 | 0.143 | 0.204 | 0.206 | 0.226 | 0.229 | 0.229 | 0.230 |

Note: The values in parentheses represent the standard errors. * , ** , and *** indicate significance at the 5%, 1%,

and 0.1% levels, respectively.

## 3.1.2. Form of the dependent variable

Since wages, as a form of income, typically exhibit a log-normal distribution, it is essential to discuss whether to present wages in logarithmic form in the model specification.

Preliminary analysis through histograms of wages and their logarithmic values suggests that the logarithmic values are more aligned with a normal distribution. To further examine whether important nonlinear relationships are overlooked between the two, this study employs the RESET (Regression Specification Error Test) for further investigation.

Figure 3-1: Histogram of Wage and Logarithmic Wage Distribution

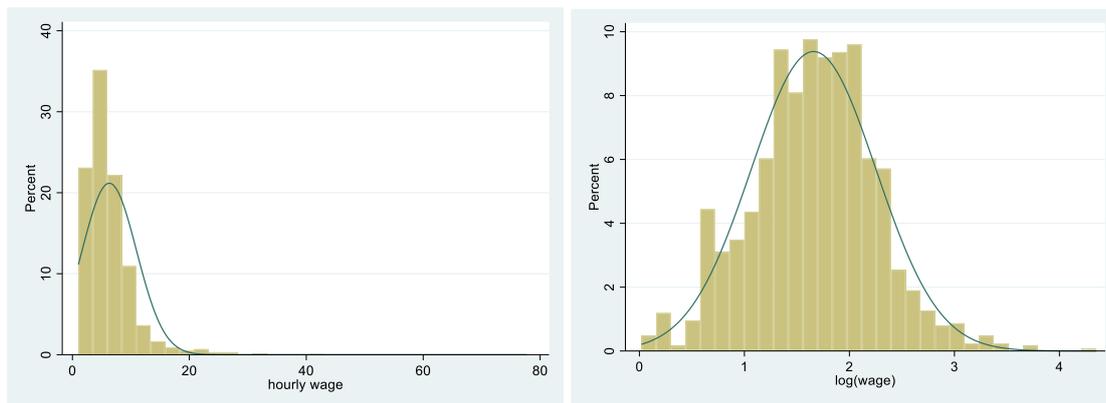

The results of the RESET test are presented in Table 2-3. The F-test statistics for both wages and logarithmic wages are not significant, indicating that neither functional form suffers from omitted variable bias. Therefore, considering the normality of data distribution, it is more reasonable to choose the logarithm of wages as the dependent variable.

Table 3-2: RESET Test Results for Wages and Logarithmic Wages

|  | (1) wage | (2) Logarithm of Wages |
|---|---|---|
| Appearance Score | 0.018 | 0.003 |
|  | (0.310) | (0.035) |
| Years of Education | -0.010 | 0.011 |
|  | (0.230) | (0.026) |
| Years of Work Experience | 0.024 | 0.007 |
|  | (0.153) | (0.017) |
| Square of Work Experience | -0.001 | -0.000 |
|  | (0.002) | (0.000) |
| Female | -0.167 | -0.113 |
|  | (1.446) | (0.164) |
| Married | -0.031 | -0.044 |
|  | (0.456) | (0.052) |
| Residing in Large Cities | -0.161 | -0.008 |
|  | (1.145) | (0.130) |
| Residing in Small Cities | 0.041 | 0.017 |
|  | (0.468) | (0.053) |

|               | (1)     | (2)      |
| ------------- | ------- | -------- |
| Association Member | 0.041   | 0.092    |
|               | (0.455) | (0.052)  |
| y_hat2        | 0.080   | 0.022    |
|               | (0.113) | (0.013)  |
| y_hat3        | 0.000   | -0.001   |
|               | (0.006) | (0.001)  |
| Constant Term | 2.648   | 0.859**  |
|               | (2.772) | (0.314)  |
| R-squared     | 0.240   | 0.401    |

Note: Standard errors are in parentheses. *, **, and *** denote significance at the 5%, 1%, and 0.1% levels, respectively.

## 3.2. Heteroscedasticity Issues

### 3.2.1. White Test

The results of the White test (see Table 2-6) indicate that neither the squared nor the cubic terms of the logarithmic wage estimates pass the significance tests, suggesting that the model does not exhibit heteroscedasticity. Therefore, without further establishing heteroscedasticity in the model, a Breusch-Pagan (BP) test was conducted.

Table 3-3: Results of the White Test

|               | (2) lu_hat2 |
| ------------- | ----------- |
| y_hat         | -0.083      |
|               | (0.116)     |
| y_hat2        | 0.001       |
|               | (0.002)     |
| Constant Term | 0.293*      |
|               | (0.126)     |
| R-squared     | 0.00        |

Note: Standard errors are in parentheses. *, **, and *** denote significance at the 5%, 1%, and 0.1% levels, respectively.

### 3.2.2. Breusch-Pagan Test

The results of the Breusch-Pagan test (see Table 3-3) indicate that the "Association Member" variable is significant, with an F-test p-value of 0.07. This suggests that the model exhibits a certain degree of heteroscedasticity.

Table 3-4: Results of the Breusch-Pagan Test

|                   | (1) lu_hat2 |
| ----------------- | ----------- |
| Appearance Score  | 0.018       |
|  | (0.017) |
| Years of Education | -0.005 |
|  | (0.005) |
| Years of Work Experience | 0.006 |
|  | (0.004) |
| Square of Work Experience | -0.000 |
|  | (0.000) |
| Female | -0.002 |
|  | (0.026) |
| Married | 0.031 |
|  | (0.028) |
| Residing in Large Cities | 0.017 |
|  | (0.033) |
| Residing in Small Cities | -0.009 |
|  | (0.027) |
| Association Member | -0.085** |
|  | (0.027) |
| Constant Term | 0.167 |
|  | (0.089) |
| R-squared | 0.013 |

Note: Standard errors are in parentheses. *, **, and *** denote significance at the 5%, 1%, and 0.1% levels, respectively.

### 3.2.3. Heteroscedasticity Correction

After correcting for heteroscedasticity using Feasible Generalized Least Squares (FGLS), the final results are presented in Table 3-6. All explanatory variables, except for the "Married" variable, show improved significance.

Table 3-5: Regression Results After Heteroscedasticity Correction（FGLS）

|  | (1) Robust | (2) FGLS |
|---|---|---|
| Appearance Score | 0.061** | 0.065*** |
|  | (0.020) | (0.019) |
| Years of Education (Years) | 0.062*** | 0.060*** |
|  | (0.005) | (0.005) |
| Years of Work Experience | 0.040*** | 0.042*** |
|  | (0.004) | (0.004) |
| Square of Work Experience | -0.001*** | -0.001*** |
|  | (0.000) | (0.000) |
| Female | -0.428*** | -0.407*** |
|  | (0.030) | (0.029) |
| Married | 0.040 | 0.049 |
|  | (0.030) | (0.030) |
| Residing in Large Cities | 0.246*** | 0.248*** |
|  | (0.038) | (0.036) |
| Residing in Small Cities | 0.103*** | 0.099** |
|  | (0.030) | (0.030) |
| Association Member | 0.174*** | 0.171*** |
|  | (0.028) | (0.027) |
| Constant Term | 0.226* |  |
|  | (0.103) |  |
| R-squared | 0.399 | 0.935 |

Note: Standard errors are in parentheses. * , **, and *** denote significance at the 5%, 1%, and 0.1% levels, respectively.

## 3.3. Multi collinearity Issues

The results of the VIF test are shown in Table 3-7. Except for the variables of work experience and its square, all other explanatory variables have VIF values less than 10. Therefore, the indicators selected in this study do not exhibit severe multicollinearity issues.

Table 3-6: Results of the VIF Test

| Variable Name | VIF | 1/VIF |
|---|---|---|
| Years of Work Experience (Years) | 16.16 | 0.0619 |
| Square of Work Experience | 15.88 | 0.0630 |
| Residing in Large Cities | 1.400 | 0.713 |
| Residing in Small Cities | 1.360 | 0.737 |
| Married | 1.200 | 0.833 |
| Female | 1.160 | 0.860 |
| Years of Education | 1.130 | 0.882 |
| Appearance Score | 1.040 | 0.957 |
| Association Member | 1.030 | 0.967 |
| Mean VIF |  | 4.490 |

## 3.4. Testing for Differences in Group Coefficients After Group Regression (Chow Test)

To examine whether there are significant differences in the coefficients of appearance on wages under different levels of education, we introduce an interaction term. A dummy variable is created, where if a person is considered attractive, it is set to 1, and 0 otherwise. The model is specified as follows:

$$\text{LWage}_i = \alpha + \gamma \cdot D_i + \beta \cdot looks + \delta(D_i \times looks) + \text{Controls}_i \cdot \gamma + +\epsilon_i$$

Similarly, we can test whether there are significant differences in the coefficients of education on wages under different levels of appearance by introducing another dummy variable. If a person has received higher education, the variable is set to 1; otherwise, it is set to 0. The model is specified as follows:

$$\text{LWage}_i = \alpha + \gamma \cdot D_i + \beta \cdot educ + \delta(D_i \times educ) + \text{Controls}_i \cdot \gamma + +\epsilon_i$$

Finally, we will examine whether there are significant differences in the coefficients of appearance on wages across different genders. The model is specified as follows:

$$\text{LWage}_i = \alpha + \gamma \cdot \text{female} + \beta \cdot \text{looks} + \delta(D_i \times \text{looks}) + \text{Controls}_i \cdot \gamma + + \epsilon_i$$

Table 3-7: Differences in Wages Based on Appearance Under Different Education Levels

|  | lwage |
|---|---|
| Is Higher Education | 0.248 |
|  | (0.130) |
| Higher Education * Appearance Score | 0.009 |
|  | (0.040) |
| Appearance Score | 0.077** |
|  | (0.026) |
| Years of Work Experience (Years) | 0.042*** |
|  | (0.004) |
| Square of Work Experience | -0.001*** |
|  | (0.000) |
| Female | -0.428*** |
|  | (0.030) |
| Married | 0.048 |
|  | (0.032) |
| Residing in Large Cities | 0.284*** |
|  | (0.038) |
| Residing in Small Cities | 0.122*** |
|  | (0.031) |
| Association Member | 0.173*** |
|  | (0.030) |
| Constant Term | 0.821*** |
|  | (0.099) |
| Sample Size | 1260 |
| $R^2$ | 0.373 |
| adj. $R^2$ | 0.368 |
| F | 74.40 |
| AIC | 1697.469 |

Note: Standard errors are in parentheses. *, **, and *** denote significance at the 5%, 1%, and 0.1% levels, respectively.

Table 3-8: Differences in Wages Based on Education Under Different Levels of Appearance

|  | (1) |
|---|---|
| Is Attractive | 0.095 |
|  | (0.203) |
| Attractiveness * Years of Education | 0.004 |
|  | (0.016) |
| Years of Education | 0.064*** |
|  | (0.013) |
| Years of Work Experience | 0.042*** |
|  | (0.007) |
| Square of Work Experience | -0.001*** |
|  | (0.000) |
| Female | -0.379*** |
|  | (0.045) |
| Married | 0.006 |
|  | (0.047) |
| Residing in Large Cities | 0.219*** |
|  | (0.057) |
| Residing in Small Cities | 0.072 |
|  | (0.046) |
| Association Member | 0.218*** |
|  | (0.049) |
| Constant Term | 0.261 |
|  | (0.172) |
| Sample Size | 538 |
| $R^2$ | 0.404 |
| adj. $R^2$ | 0.392 |
| F |  |
| AIC | 715.985 |

Note: Standard errors are in parentheses. *, **, and *** denote significance at the 5%, 1%, and 0.1% levels, respectively.

Table 3-9: Differences in Wages Based on Gender Under Different Levels of Appearance

|  | lwage |
|---|---|
| Is Female | -0.724*** |
|  | (0.103) |
| Female * Appearance Score | 0.092** |
|  | (0.031) |
| Years of Education | 0.063*** |
|  | (0.005) |
| Years of Work Experience (Years) | 0.040*** |
|  | (0.004) |
| Square of Work Experience | -0.001*** |
|  | (0.000) |
| Married | 0.040 |
|  | (0.031) |
| Residing in Large Cities | 0.243*** |
|  | (0.037) |
| Residing in Small Cities | 0.099** |
|  | (0.030) |
| Association Member | 0.175*** |
|  | (0.030) |
| Constant Term | 0.408*** |
|  | (0.080) |
| Sample Size | 1260 |
| R2 | 0.398 |
| adj. R2 | 0.394 |
| F | 91.88 |
| AIC | 1644.516 |

Note: Standard errors are in parentheses. * , **, and *** denote significance at the 5%, 1%, and 0.1% levels, respectively.

Furthermore, the study analyzes the impact of appearance based on educational attainment and draws the following conclusions: (1) There is evidence of appearance-based discrimination in the low-education group; however, in the high-education group, neither a "beauty premium" nor an "ugliness penalty" is observed. Therefore, transitioning from a low-education group to a high-education group can partially mitigate the "ugliness penalty." At least among workers with high education, lower "aesthetic appeal" does not significantly affect income. Consequently, obtaining a higher education can be regarded as a means to compensate for appearance disadvantages and reduce the "ugliness penalty."

The study further examines the impact of education based on appearance scores and reaches the following conclusions: (2) Education does not provide a "degree premium" for individuals with attractive appearance, nor does it incur an "educational penalty." Therefore, regardless of a person's attractiveness, pursuing higher education remains an important pathway to increasing wages.

Additionally, the analysis of the role of appearance by gender leads to the conclusion: (3) The appearance advantage for women is more pronounced than that for men. Thus, for women, improving their educational attainment translates to a higher "beauty premium" (data-driven perspective, not a personal viewpoint). This indicates that women should invest more time in the process of enhancing their education.

# 4.Conclusion

The empirical research outlined above indicates that the appearance-wage effect among Chinese workers is highly significant. Ceteris paribus, possessing an appearance advantage can substantially increase wage income due to the "beauty premium," while a less favorable appearance can significantly decrease wages, resulting in an "ugliness penalty." This conclusion holds true across the entire sample as well as within male and female subgroups. In other words, regardless of gender, the level of physical appearance significantly affects wage levels. The prevailing notion that appearance is not important for men is unfounded; like women, attractive men also enjoy higher wages, while unattractive men face lower wages.

Future research could incorporate deep learning models [18, 19] such as Transformers [20], Mamba [21], to construct corresponding appearance-wage impact models. By integrating efficient attention mechanisms, model performance could be further enhanced [22], allowing for a deeper investigation into the issue of appearance discrimination.

In light of these findings on appearance discrimination, the concept of "dress for success" (Daniel S. Hamermesh et al., 1999) underscores the necessity for workers to pay attention to their personal image. Additionally, since the negative effect of appearance disadvantage on wage income is not significant among highly educated individuals but is pronounced among those with lower educational attainment, enhancing educational levels is an effective strategy to mitigate income reductions linked to appearance. Finally, as there is neither a "beauty premium" nor an "ugliness penalty" among highly educated groups, the government could reduce appearance discrimination in the labor market by facilitating greater access to higher education for workers.

# 5.References

[1] Wu, D. (2016). The Economics of Appearance. Discovery, (1), 8-9.

[2] 2018 China Youth Appearance Competitiveness Report: Women at 18, Men at 40 Enter the Beauty Competition [Z/OL]. [2018-05-09].

[3] Kuhn, P., & Shen, K. (2011). Gender Discrimination in Job Ads: Theory and Evidence. NBER Working Paper No. 17453,2011.

[4] Zhong, W. (2006). The Economic Benefits of Beauty Compared to Doctoral Degrees. Elite Talent, (3), 24.

[5] Hamermesh，D S and Biddle，J E. Beauty and the Labor Market ［J］. The American Economic Review，1994，84：1174-1194.

[6] Hamermesh，D S，Meng，X and Zhang，J S. Dress for Success—Does Primping Pay? ［Z］. NBER Working Paper，No.W7167，1999.

[7] Hamermesh D S. Beauty Pays：Why Attractive People are More Successful ［M］. Princeton University Press，2011.

[8] Quinn，R. E. ，1978，"Productivity and the process of organization improvement: Why we cannot talk to each other"，Public Administration Review. 38:41 －45.

[9] Marlowe C M，Schneider S L，Nelson C E. Gender and Attractiveness Biases in Hiring Decisions：Are More Experienced Managers Less Biased?［J］. Journal of Applied Psychology，1996，81（1）：11-21.

[10] Hosoda M，Stone-Romero E F，Coats G. The Effects of Physical Attractiveness on Job-Related Outcomes：A Meta-Analysis of Experimental Study ［J］. Personnel Psychology，2003，56（2）：431-462.

[11] Zhang, J. L. (2004). Does Appearance Affect Interview Outcomes? Occupation, (5), 31-32.

[12] Gao, W. S. (2009). Investing in Healthy Human Capital, Height, and Wage Returns: An Empirical Study of Household Survey Data from 12 Cities. Chinese Population Science, (3), 76-85.

[13] Liu, Y. P., Zheng, Y., & Zhang, C. C. (2016). Does Being Attractive Lead to High Income? A Study of Appearance Discrimination in the Chinese Labor Market. Economic Review, (5), 83-95.

[14] Wang, X., Yue, Y., & Zhu, C. (2018). Higher Appearance, Higher Income? An Empirical Study Based on the 2014 "China Labor Dynamics Survey". Journal of Yunnan University of Finance and Economics, (5), 79-91.

[15] Guo, J. Q., Fei, S. L., & Lin, P. (2017). Is Higher Beauty Linked to Higher Income? Discussing the "High Heel Curve" of Appearance and Income. Economics (Quarterly), (1), 147-172.

[16] Yuan, H. J., & Zhang, Z. Y. (2005). The Expression of Implicit Appearance Bias in Recruitment Contexts. Journal of Peking University (Natural Science Edition), 41(2), 303-308.

[17] Yang, Y. Z., Fang, X. M., & Zheng, X. D. (2017). The Educational Corrective Effect of Appearance Discrimination in the Labor Market. Southern Economy, (3), 71-98.

[18] Wang, R.F., & Su, W.H. (2024). The Application of Deep Learning in the Whole Potato Production Chain: A Comprehensive Review. *Agriculture, 14*(8), 1225.

[19] Yao, M., Huo, Y., Ran, Y., Tian, Q., Wang, R., & Wang, H. (2024). Neural Radiance Field-based Visual Rendering: A Comprehensive Review. *arXiv preprint arXiv:2404.00714*.

[20] Wang, Z., Wang, R., Wang, M., Lai, T., & Zhang, M. (2024). Self-supervised transformer-based pre-training method with General Plant Infection dataset. *arXiv preprint arXiv:2407.14911*.

[21] Yao, M., Huo, Y., Tian, Q., Zhao, J., Liu, X., Wang, R., & Wang, H. (2024). FMRFT: Fusion Mamba and DETR for Query Time Sequence Intersection Fish Tracking. *arXiv preprint arXiv:2409.01148*.

[22] Huo, Y., Yao, M., Tian, Q., Wang, T., Wang, R., & Wang, H. (2024). FA-YOLO: Research On Efficient Feature Selection YOLO Improved Algorithm Based On FMDS and AGMF Modules. *arXiv preprint arXiv:2408.16313*.